\documentclass{ws-ijmpe}

\begin{document}

\markboth{ L. M. Robledo, M. Warda}{Cluster radioactivity of Th isotopes in the mean-field HFB theory.}

\catchline{}{}{}{}{}
%
%%%%%%%%%%%%%%%%%%%%%%%%%%%%%%%%%%%%%%%%%%%%%%%%%%%%%%%%%%%%%%%%%%%%

\title{Cluster radioactivity of Th isotopes in the mean-field HFB theory.}

\author{L. M. ROBLEDO\textsuperscript{a}, M. WARDA\textsuperscript{b,c}}

\address{\textsuperscript{a} Departamento de F\'\i sica Te\'orica C-XI, 
	Universidad Aut\'onoma de Madrid, \\
        28-049  Madrid, Spain\\
%	\\
	\textsuperscript{b} Departament d'Estructura i Constituents de la Mat\`eria
	and Institut de Ci\`encies del Cosmos, \\
	Facultat de F\'{\i}sica, Universitat de Barcelona,
	Diagonal {\sl 647}, {\sl 08028} Barcelona, Spain\\
%	\\
	\textsuperscript{c} Katedra Fizyki Teoretycznej, Uniwersytet Marii Curie--Sk\l odowskiej,\\
       ul. Radziszewskiego 10, 20-031 Lublin, Poland\\
	warda@kft.umcs.lublin.pl}  
%}

\maketitle

\begin{history}
\received{(received date)}
\revised{(revised date)}
%\accepted{(Day Month Year)}
%\comby{(xxxxxxxxxx)}
\end{history}

\begin{abstract}

Cluster radioactivity is described as a very mass asymmetric fission process. 
The reflection symmetry breaking octupole moment has been
used in a mean field HFB theory as  leading coordinate instead of the  
quadrupole moment usually used in standard fission calculations. The procedure has been
applied to the study of the ``very mass asymmetric fission barrier'' of several
even-even Thorium isotopes. The masses of the emitted clusters as well as the
corresponding half-lives have been evaluated on those cases where  experimental
data exist.

\end{abstract}

\section{Introduction}

Cluster radioactivity, first predicted theoretically by Sandulescu et
al.\cite{1} in 1980,  was discovered in 1984 by Rose and Jones\cite{2} in the
spontaneous reaction $^{223}$Ra~$\rightarrow ^{14}$C$+^{209}$Pb.  Although
cluster emission is a very exotic process, with a relative branching ratio to
$\alpha$-decay of the order of $10^{-10}$ -- $10^{-17}$, it has been observed in
many actinide nuclei from $^{221}$Fr to $^{242}$Cm. Clusters emitted in these
reactions are light nuclei from $^{14}$C to $^{34}$Si, whereas the heavy mass
residue is a nucleus that  differs from the doubly magic $^{208}$Pb by no more
than four nucleons.

Cluster emission fills up the gap in the fragment's mass spectrum of the
nuclear decay between $\alpha$ radioactivity and spontaneous fission, where the
masses of the fragments are typically greater than 60. From a theoretical point of view,
cluster radioactivity may be treated as the emission of a pre-formed cluster inside
the nucleus in close analogy to $\alpha$-decay. The alternative approach is to
consider these reactions as a particular case of very mass asymmetric fission. 

In this paper the potential energy surfaces (PES) of several even mass Th isotopes 
obtained with the help of the
Hartree-Fock-Bogoliubov (HFB) theory and the D1S Gogny force are analyzed. As it is 
well established,  this methodology and force have been successfully applied in the calculation of the
spontaneous fission properties of heavy nuclei.\cite{3,4,5} Therefore, it seems natural
to think that this method
could also be applied  to investigate very asymmetric fission leading to the emission
of clusters. Preliminary explorations in this direction\cite{6} have shown that this is indeed the case
and with the present calculation we want to extend the description to other nearby region of the
Nuclide Chart.

\section{Theoretical Model and Results}

Fission barriers  are obtained in the mean-field models by calculating 
potential energy curves as a function of a convenient quantity by considering a
constraint applied on the system. Usually a single constraint on some elongation
parameter e.g. quadrupole moment is used and separate fission paths are
obtained. This procedure sometimes may lead to rather incomplete or even
misleading conclusions about the topology of the fission barrier\cite{4} and
therefore calculations with simultaneous constraints on quadrupole and octupole
moments have been performed in our case. As a consequence a bidimensional PES
has been created as a function of both the elongation and reflection-asymmetry
parameters of the nuclear system. Afterwards, fission paths have been found in
the bottom of the valleys of the surface. Such procedure ensures that all
fission paths and the passes connecting them are properly described.

In Fig. 1 we have plotted, as an example, the PES of $^{230}$Th as a function of
quadrupole and octupole moments. The part of the curve below the dotted line
represents compact solutions, whereas the part above this line corresponds to
the system made of two separated fragments. Two valleys leading from the ground
state to scission can be easily found on the surface. The fission paths, which
are the bottoms of these valleys are marked with thin solid lines. One of them
goes initially  along $Q_3=0$ axis and relatively small reflection asymmetry can
be found from around $Q_2=50$ b. This valley leads to normal spontaneous
fission.  On the other valley, the octupole moment is all the time different
from zero  and increases almost linearly with quadrupole moment. The very big 
asymmetry of masses of the fragments suggests that fission along this
path leads to cluster radioactivity. 

\begin{figure}[th]
\centerline{\psfig{file=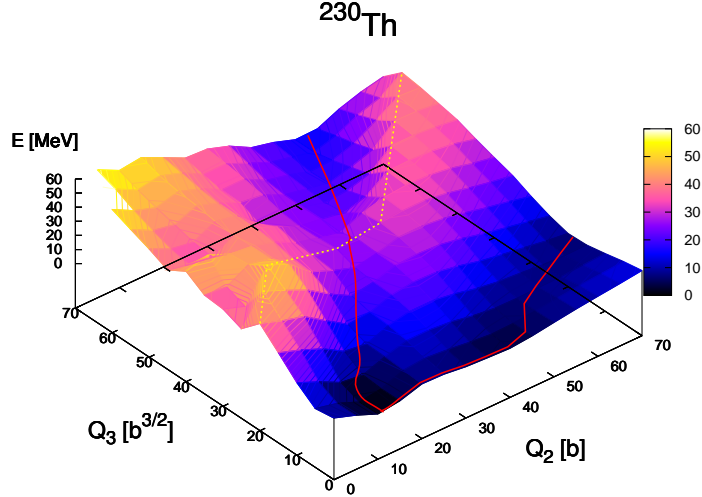,width=7cm,angle=270}}
\vspace*{8pt}
\caption{The PES of $^{230}$Th as a function of quadrupole  and octupole
moments. Thick solid lines show possible fission paths. Scission points is
shown by a dotted line. }
\end{figure}

The very mass asymmetric fission path could, in principle, be characterized and
obtained by the constraint on the quadrupole moment but it turns out in this and
other\cite{6,7} examples that using a single constraint in the octupole moment
$Q_3$ suffices to follow that path and therefore it is the octupole and not the
quadrupole moment the ``natural'' coordinate to be used in this kind of studies.
The similar conclusions can be deduced from Fig. 2, where the sequence of the
density distribution plots of $^{230}$Th shows the evolution of the shape of the
nucleus at the cluster emission path with increasing octupole moment.  One can
clearly see there that increasing $Q_3$ leads straightforwardly to very mass
asymmetric fission.

\begin{figure}[th]
\centerline{\parbox[c]{5in}{
\psfig{file=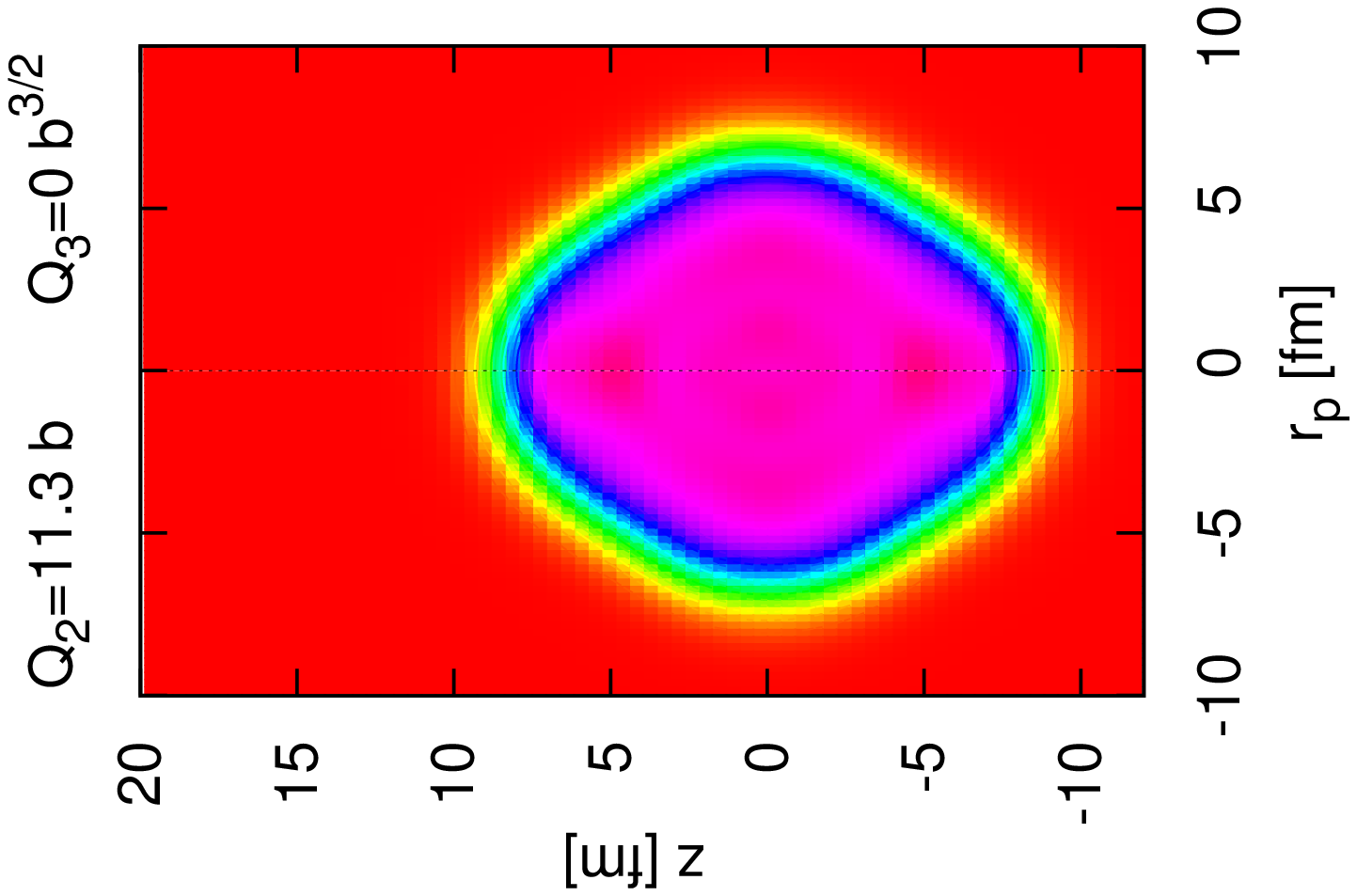,width=4.8cm,angle=270}\psfig{file=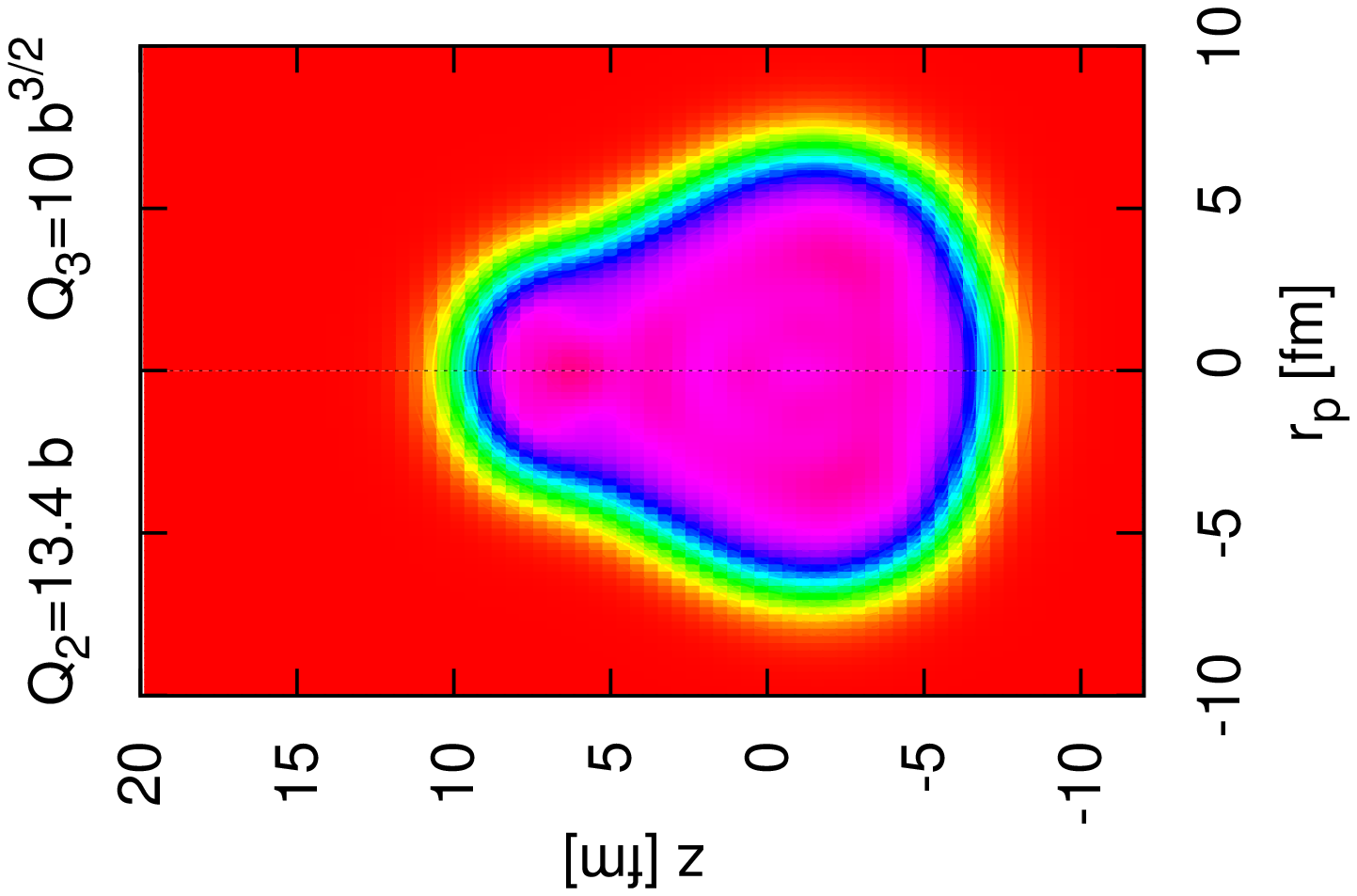,width=4.8cm,angle=270}\psfig{file=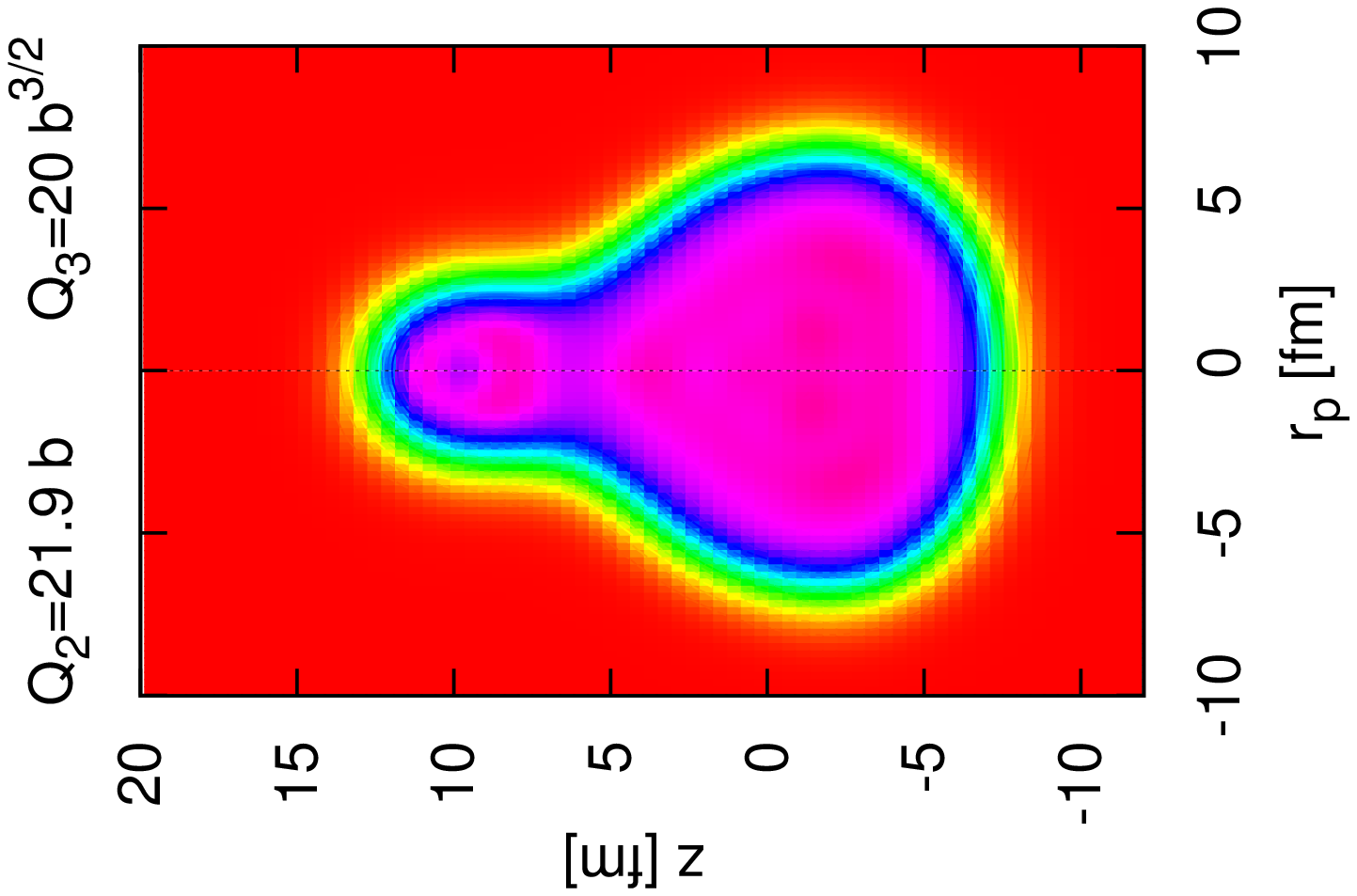,width=4.8cm,angle=270}\psfig{file=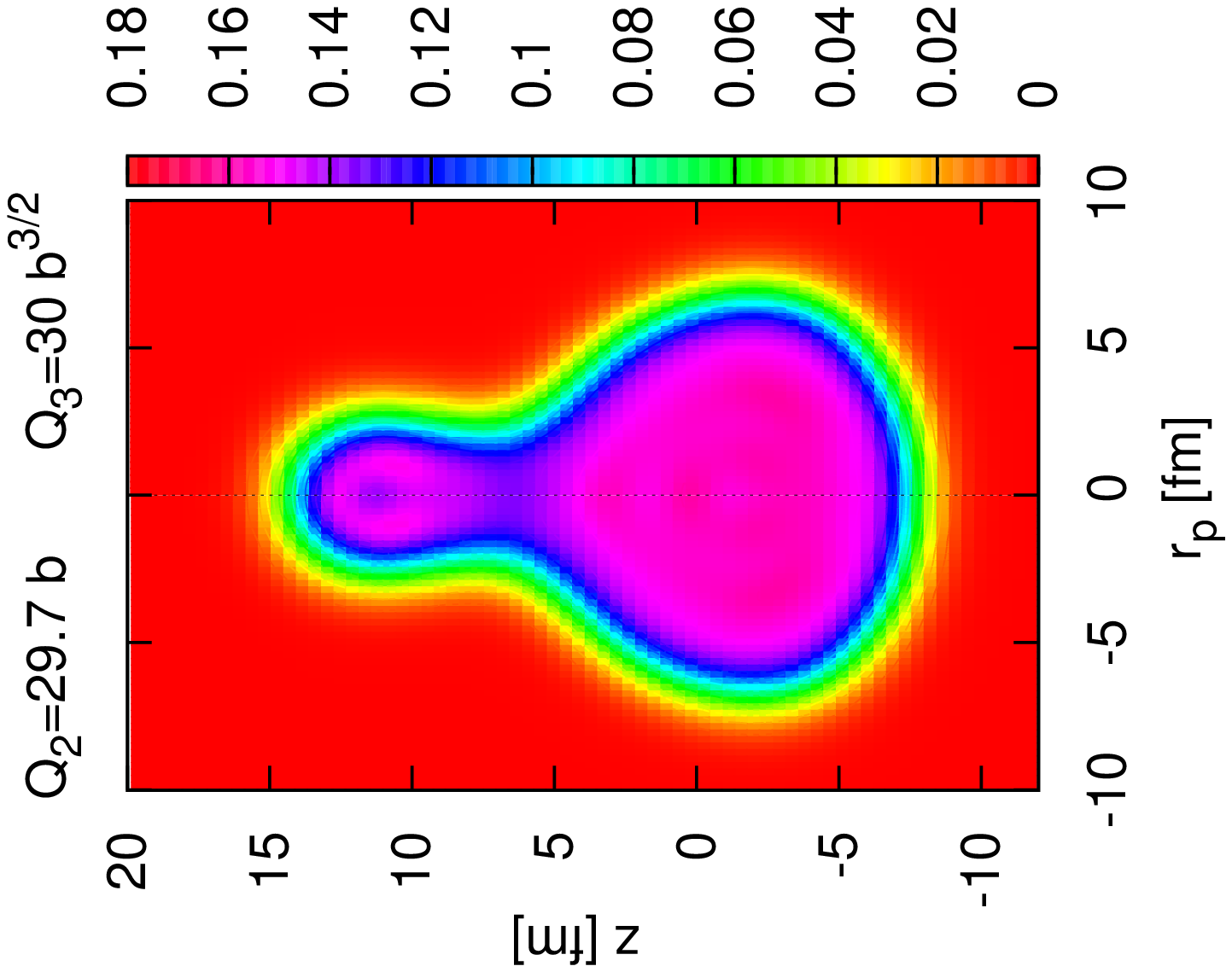,width=4.8cm,angle=270}\\
\psfig{file=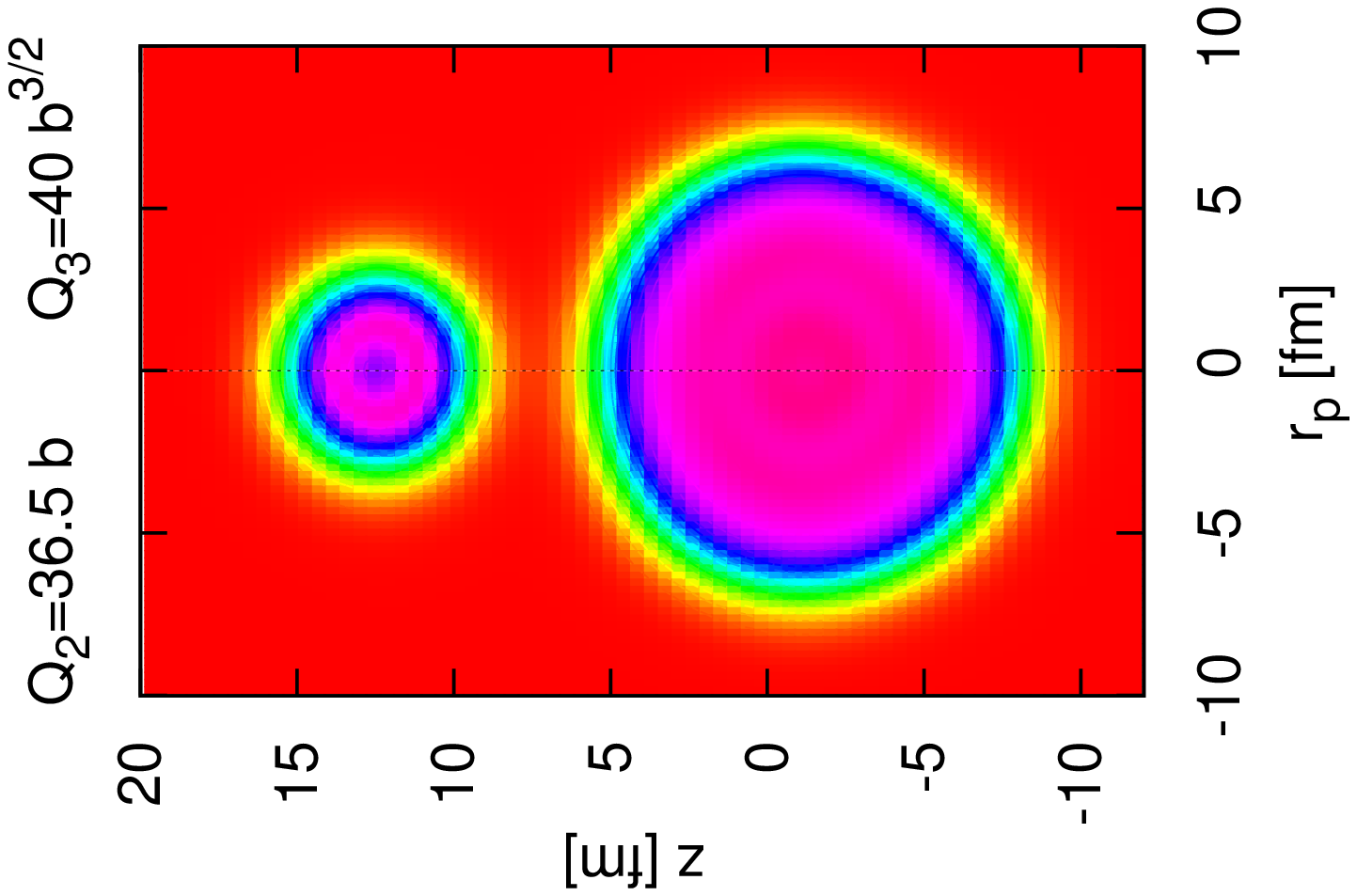,width=4.8cm,angle=270}\psfig{file=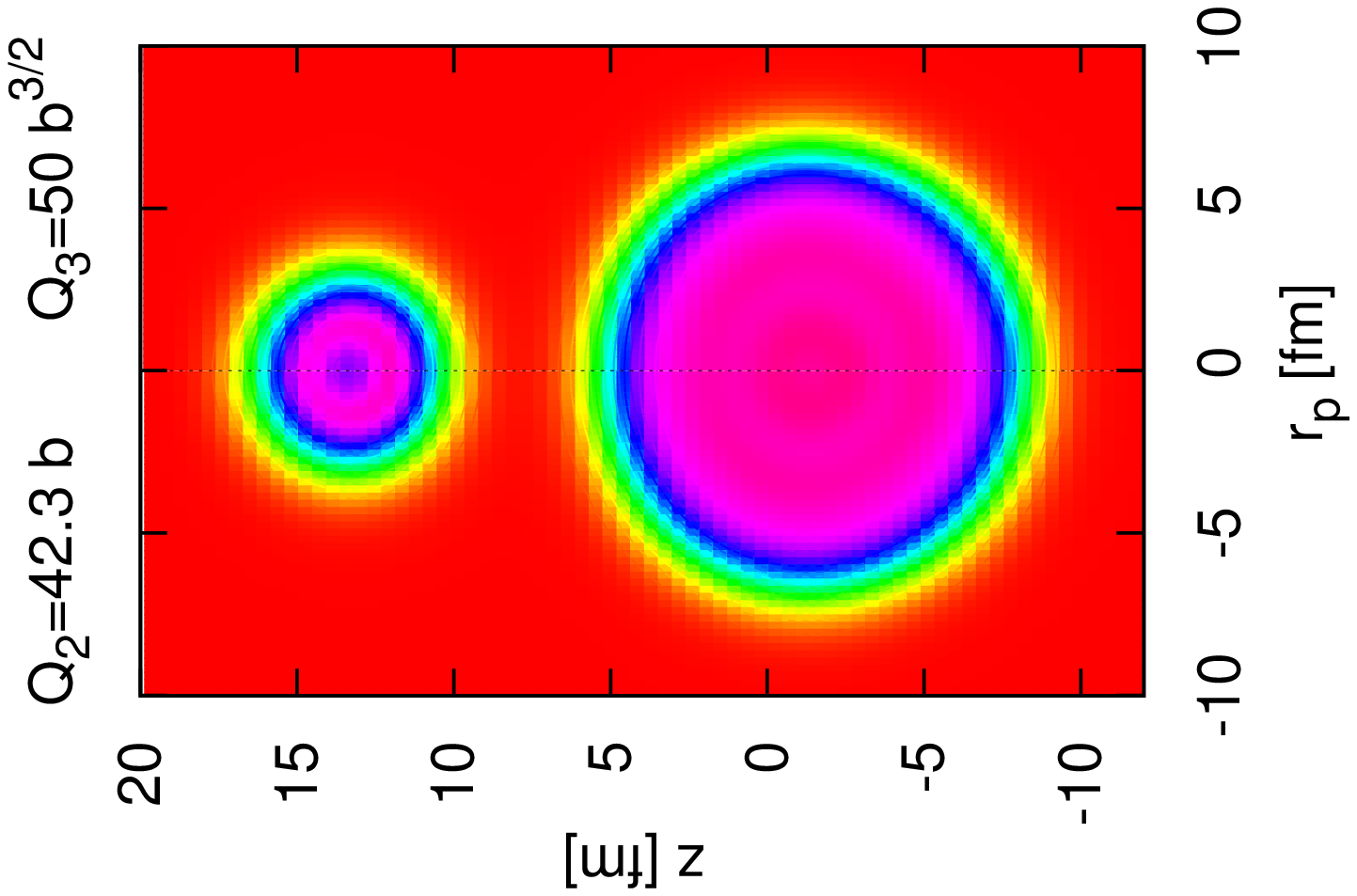,width=4.8cm,angle=270}\psfig{file=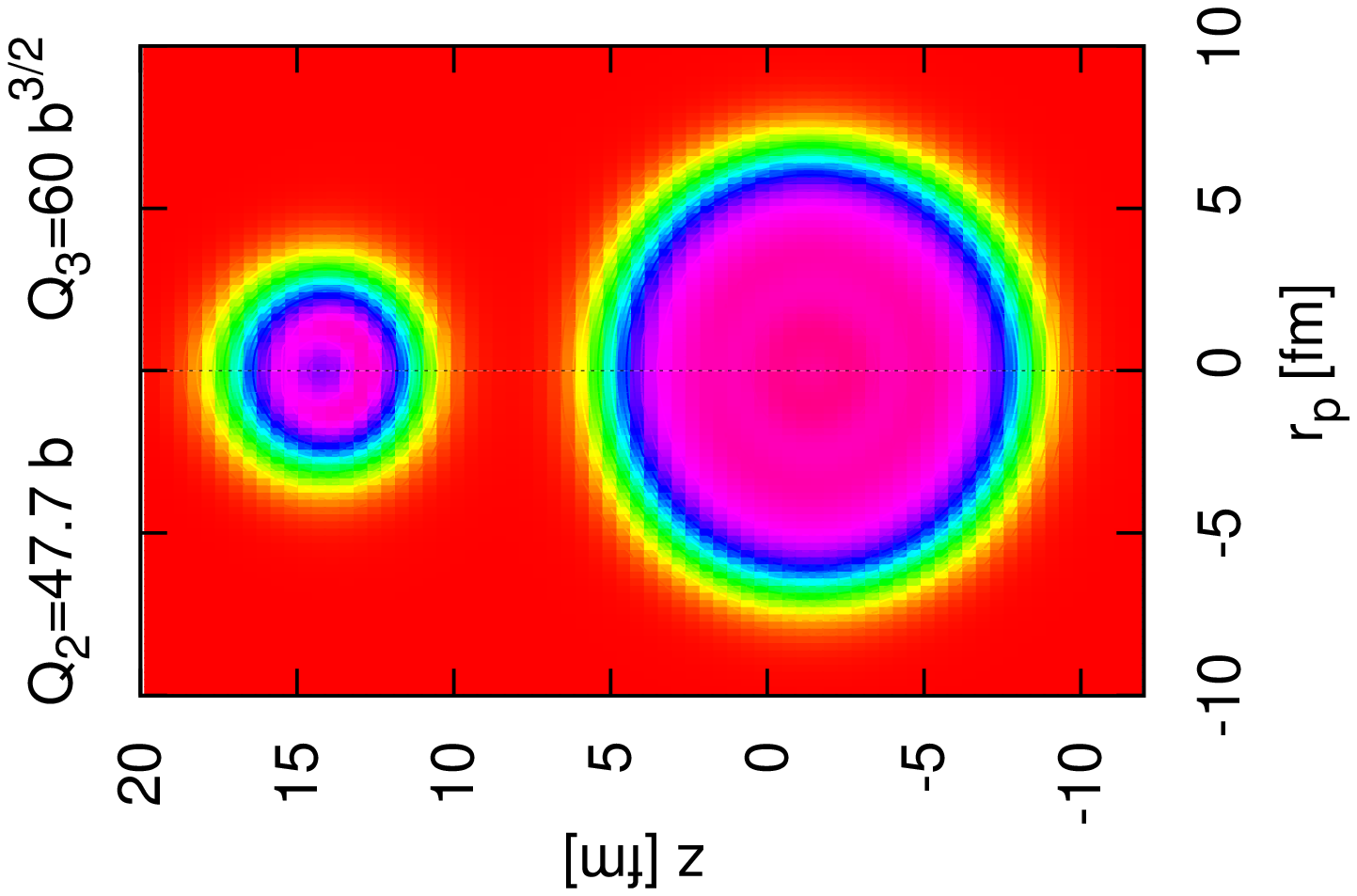,width=4.8cm,angle=270}\psfig{file=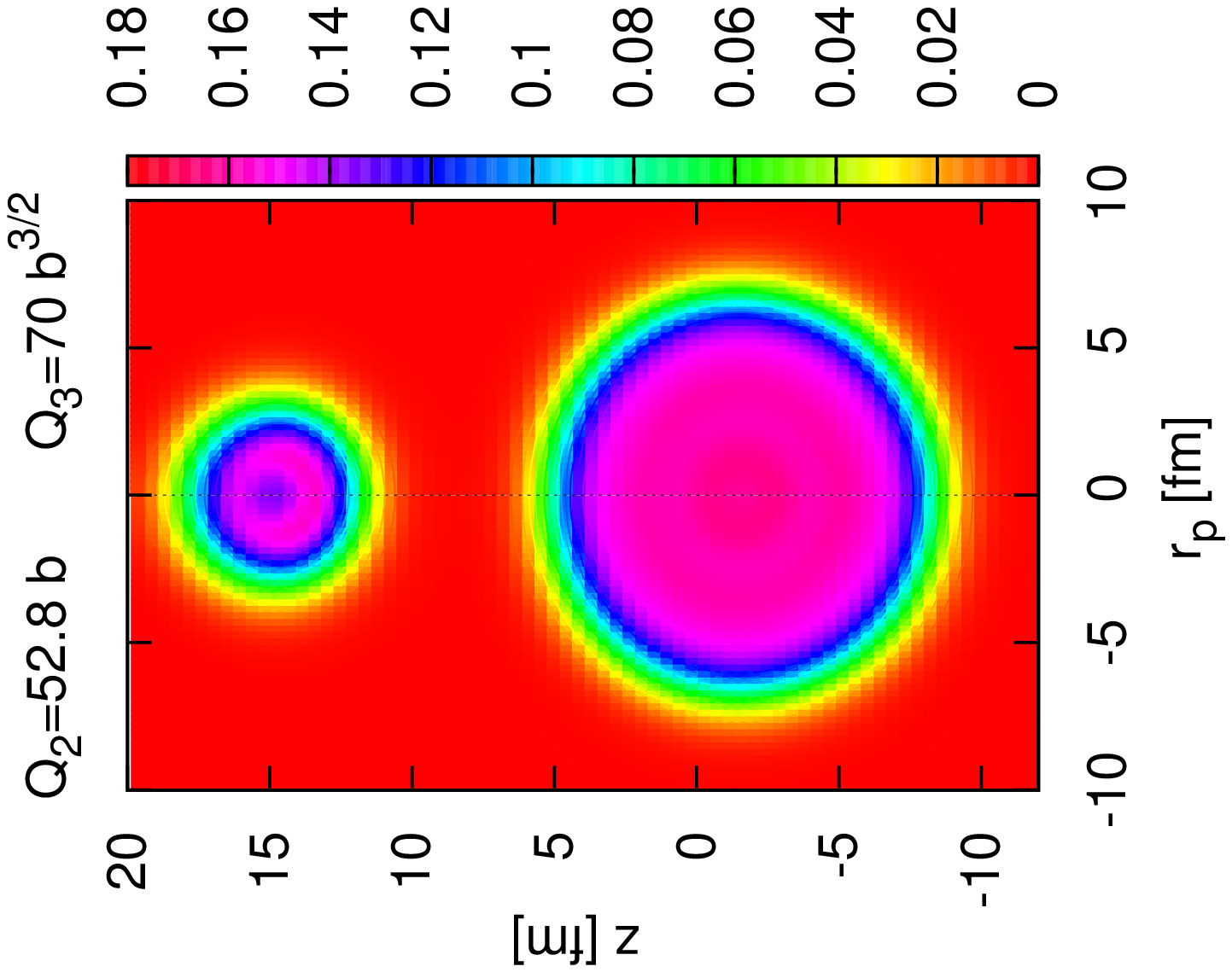,width=4.8cm,angle=270}
}}
\vspace*{8pt}
\caption{Shape evolution of $^{230}$Th along the cluster emission path.
Nuclear density distribution of nucleus is plotted for octupole moments every 10 b$^{3/2}.$}
\end{figure}

The shape of the cluster emission barriers of the Th nuclei as a function of
$Q_3$ is plotted in the lower panels of Fig. 3. The barriers are around 25 MeV
high and these  are huge values in comparison with spontaneous fission. The
barriers consist typically of two parts: the first  goes from the ground state
(which is octupole deformed in $^{226}$Th and $^{228}$Th) to the saddle point.
The scission point is localized at the top of the barrier. The second part of
the barriers, with decreasing energy as a function of the octupole moment,
corresponds  to the solution with two separated fragments. In fact, there is
small discontinuity in the fission path calculated  in our model  between the
up-going compact part and the decreasing after-scission part. This discontinuity
could be resolved and studied in detail  by using another  constraint as for
example on the neck parameter,\cite{3} the "slice" operator\cite{6} or
hexadecapole moment,\cite{4} but the corrections obtained by taking it into
account would probably have a negligible  impact on the values of the heights of
the fission barriers as well as on the half-lives.

\begin{figure}[th]
\centerline{\parbox[c]{5in}{
\psfig{file=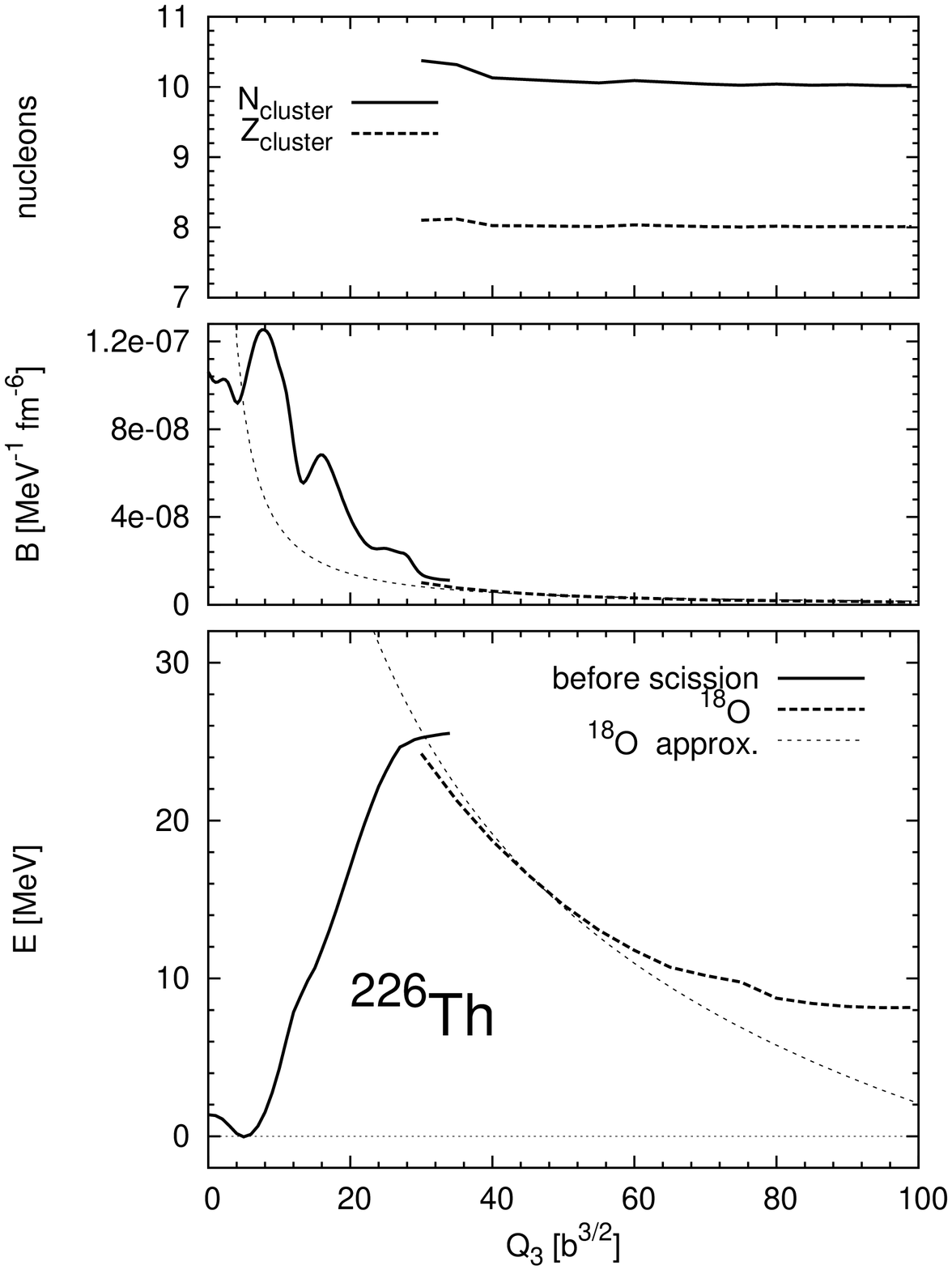,width=6cm}
\psfig{file=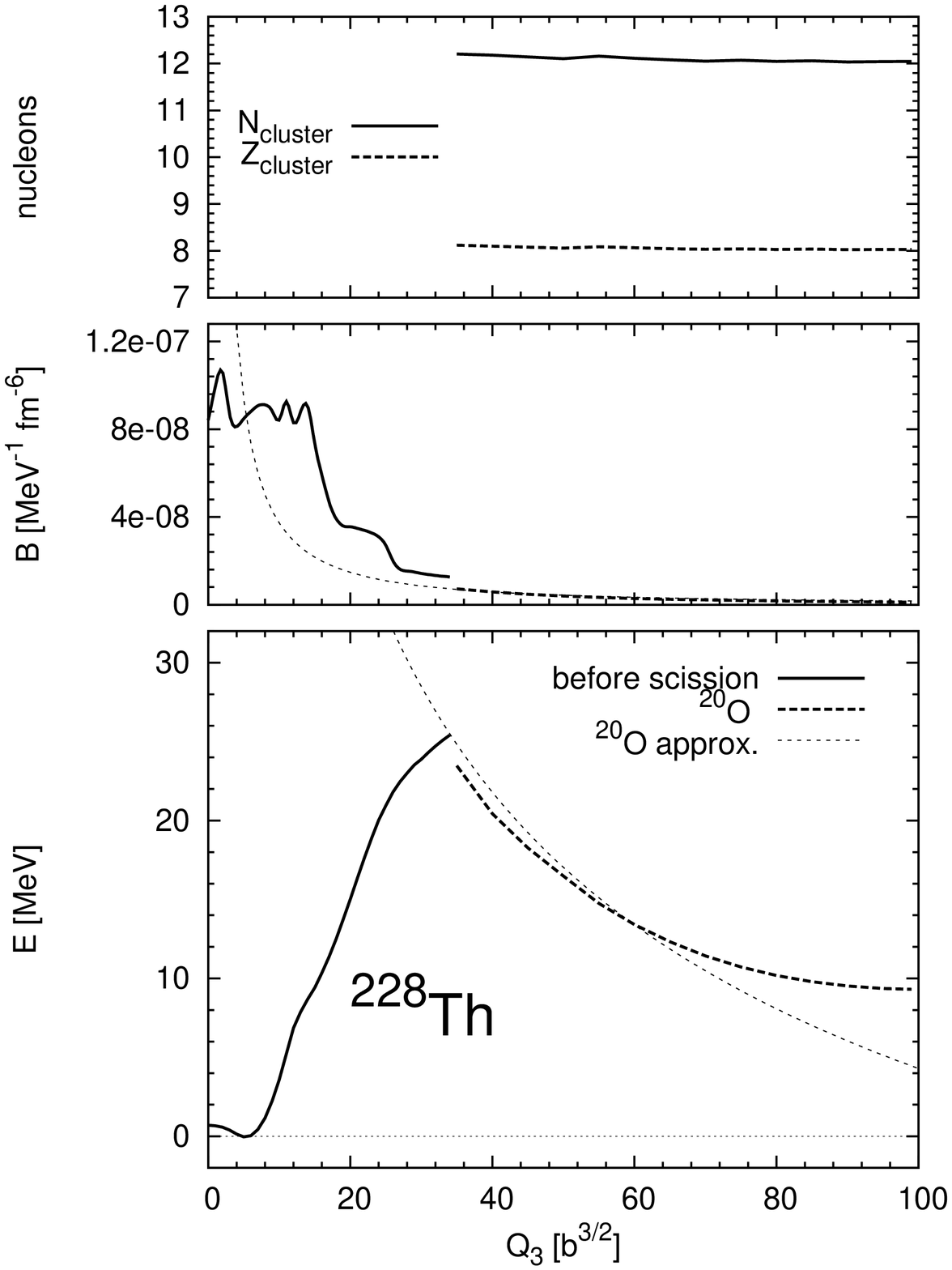,width=6cm}\\
\psfig{file=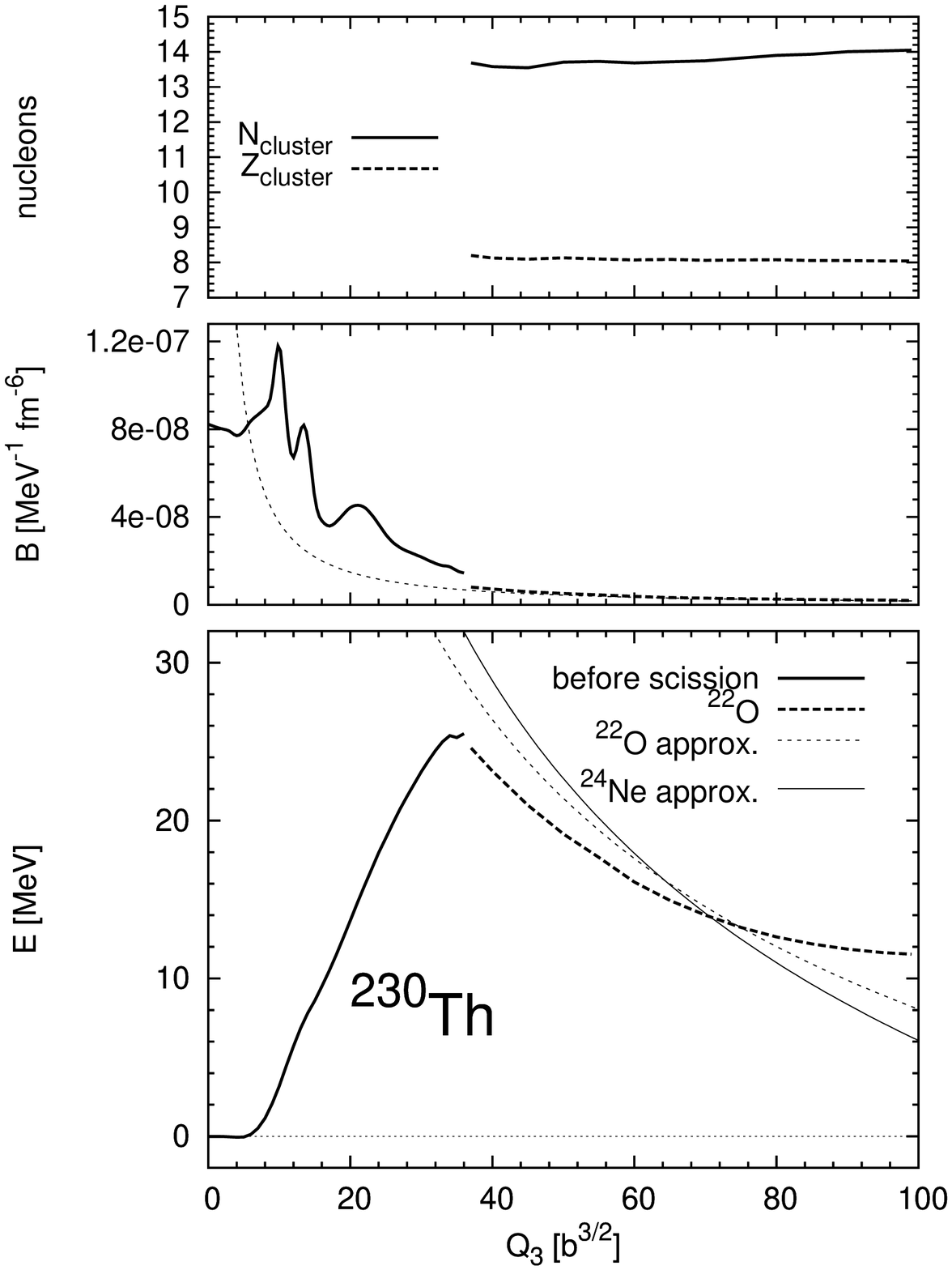,width=6cm}
\psfig{file=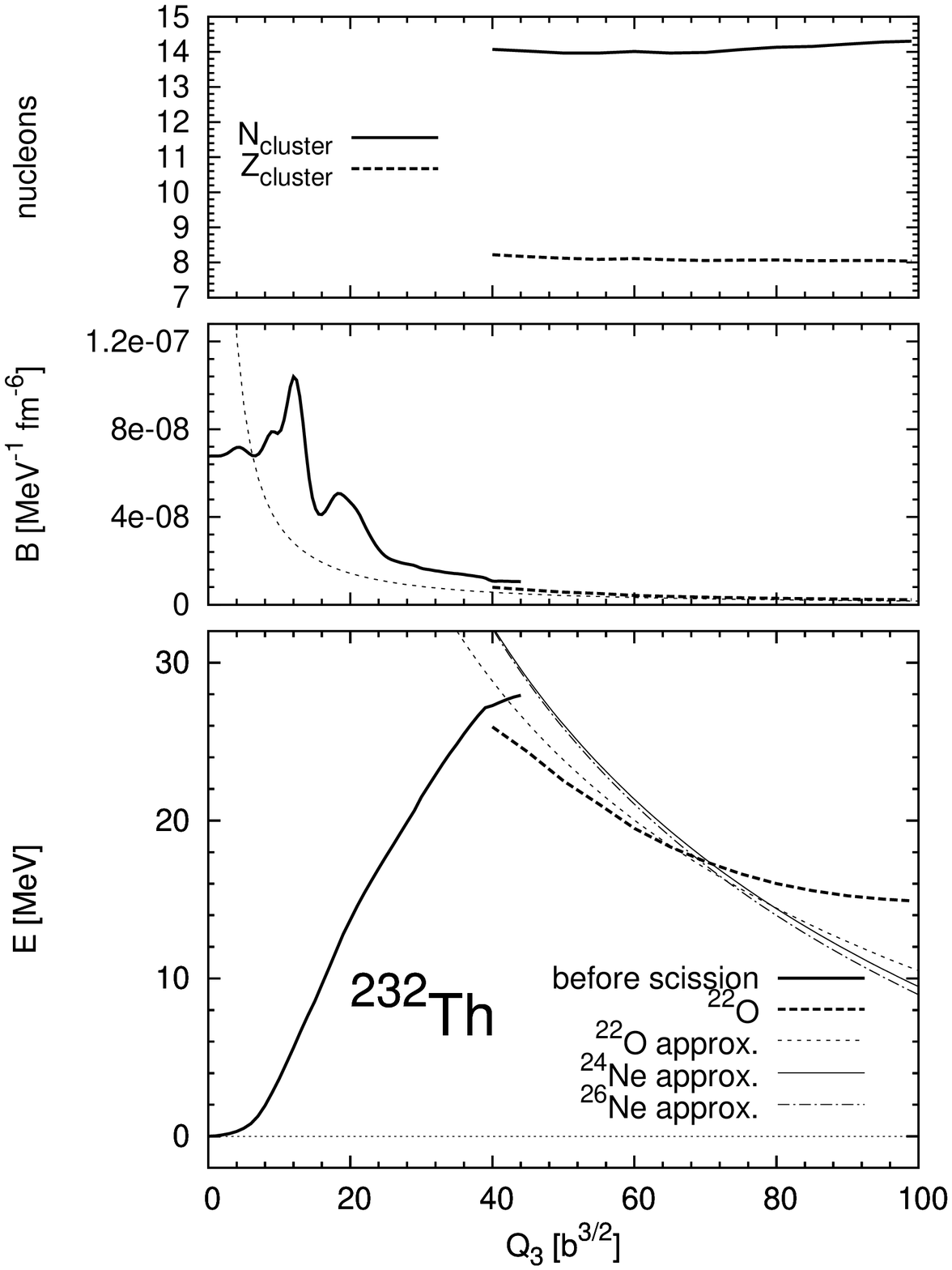,width=6cm}
}}
\vspace*{8pt}
\caption{Fission paths of $^{226}$Th, $^{228}$Th, $^{230}$Th and $^{232}$Th 
isotopes. Potential energy (bottom panels), inertia parameter $B(Q_{3})$
(middle panels) and the number of
nucleons in light fragment after scission (top panels) are plotted as
 a function of $Q_{3}$. Approximate values of the energy Eq. (1) and collective inertia
 Eq. (7) are marked with  thin lines.}
\end{figure}

The potential energy in the two-fragment branch of the fission path decreases
with increasing octupole moment mainly due to the decrease of the Coulomb
repulsion between the two outgoing fragments. This should be manifested for
large $Q_3$ values (large separation between fragments), where the nuclear
interaction between fragments is no longer relevant. To check this assumption we
will approximate the energy by the expression
		 \begin{equation}
		 V_{two\;fragments}(Q_{3})=E_0-Q +V_{Coul}(Q_{3}),
		 \end{equation}
where $E_0$ is the ground state energy calculated in the HFB theory and the Q value is obtained
from experimental binding energies.\cite{8} The Coulomb energy can be expressed as a
function of $Q_{3}$ 
		\begin{equation}
		V_{Coul}(Q_{3})=e^{2}\frac{Z_{1}Z_{2}}{R(Q_{3})},
		\end{equation} 
by means of a relation between the octupole moment and the distance between the centers of mass of the fragments. By assuming that the fragments are spherical and with a constant density we obtain
		\begin{equation}
		Q_{3}=f_{3}R^{3},
		\end{equation}
with	
		\begin{equation}
		f_{3}=\frac{A_{1}A_{2}}{A}\frac{(A_{1}-A_{2})}{A}.
		\end{equation}  
		
The potential energy from Eq. (1) is plotted in Fig. 3 with a thin line in the
PES panel. We observe quite large energy differences between this formula and the HFB results
specially at large values of $Q_3$. The differences can be attributed to slight
deviations of the fragment's density distribution from sphericity but first of
all to a not big enough size of the harmonic
oscillator basis used \cite{7}. For small $Q_{3}$ values, around the scission
point, the HFB energy looks closer to the values  of Eq. (1) as can be  seen in
Fig. 3. This is a consequence of a smaller spatial size of the system that
requires a smaller basis size for its description and also to the fact that the
nuclear interaction between fragments probably is not very relevant in that
case.

In order to calculate half-lives of cluster emission the WKB
approximation has been used.\cite{7} The half life is given by
		\begin{equation}
		t_{1/2}=2.86\,10^{-21}(1+\exp(2S)),
		\end{equation}
where $S$ is the action along the $Q_{3}$ constrained path 
		\begin{equation}
		S=\int_{a}^{b}dQ_{3}\sqrt{2B(Q_{3})(V(Q_{3})-E_{0})}.
		\end{equation}
Here $a$ and $b$ are turning points at the ground state energy $E_0$ and
$B(Q_{3})$ is  collective quadrupole inertia (computed with the standard
approximation of neglecting the residual interaction in its evaluation) with $Q_{3}$ as the collective variable. The values of $B(Q_{3})$
are plotted in the middle panels of Fig. 3. Again, for the branch of the barrier
after scission we have used  the approximate formula
		\begin{equation}
		B(Q_{3})=\frac{\mu}{9Q_{3}^{4/3}f_{3}^{2/3}},
		\end{equation}
with effective mass:
		\begin{equation}
		\mu=m_{n}\frac{A_{1}A_{2}}{A_{1}+A_{2}}.
		\end{equation}
obtained by assuming that the mass is the reduced mass of two spherical
fragments and expressed in terms of the octupole moment. The oscillations of
$B(Q_{3})$ observed in the part of the diagram before scission are due to the
shell effects in the deforming nuclei. The collective mass after scission is
smooth as the fragments almost do not change their deformation. As can be
checked in Fig. 3 the approximate expression of Eq. (7) gives here results
similar to the HFB calculations.

The half-lives of cluster emission of Th isotopes are presented in Fig. 4 and
Table~1. Good agreement is found between the theoretical results and the
experimental data. The differences do not exceed two orders of magnitude. This
could be considered as a rather poor agreement but it is to be noted that this
kind of errors is typical for other theoretical predictions in the field of
fission. It is also  worth to stress here that in our model there are no free
parameters to be fitted, making the degree of agreement between our results and
experiment quite outstanding.

\begin{figure}[th]
\centerline{\psfig{file=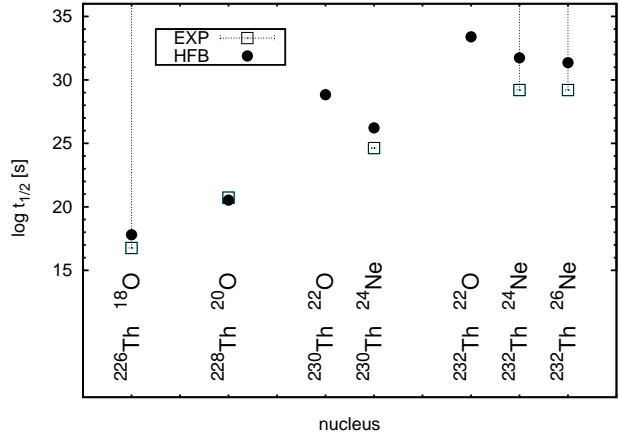,width=6cm,angle=270}}
\vspace*{8pt}
\caption{Half-lives of cluster emission of Th isotopes calculated in the HFB theory
are compared with experimental data.\protect\cite{9}}
\end{figure}

\begin{table}[pt]
\tbl{Half-lives of cluster emission of Th isotopes calculated in the HFB theory
are compared with experimental data.\protect\cite{9}}
{\begin{tabular}{@{}cccc@{}} \toprule
Emitter&Cluster &$log(t_{1/2}[s])$&$log(t_{1/2}[s])$\\ 
 & &HFB&EXP\\ 
\colrule  
$^{226}$Th& $^{18}$O &17.80&$>$16.76\\    
$^{228}$Th& $^{20}$O &20.53&20.73\\    
$^{230}$Th& $^{22}$O &28.83&--\\    
$^{230}$Th& $^{24}$Ne&26.22&24.63\\    
$^{232}$Th& $^{22}$O &33.39&--\\    
$^{232}$Th& $^{24}$Ne&31.74&$>$29.20\\    
$^{232}$Th& $^{26}$Ne&31.36&$>$29.20\\    
\botrule
\end{tabular}}
\end{table}

In  $^{230}$Th and $^{232}$Th the predicted  masses of the fragments  in the HFB
model differ slightly, by two protons, from experimental data. In both nuclei
the HFB fission path corresponds to a $^{22}$O fragment. 
Nevertheless, the
half-lives for cluster emission of Na isotopes are shorter than those found at
the bottom of the HFB fission valley. What is possibly happening in this case is that
the fission path leading at scission to a Na cluster has a lower action integral than the
fission path located at the fission valley (as a consequence of a lower collective mass).
As a consequence the Na cluster path yields a lower half-life making it the preferred decay mode.
To explore this possibility a study in terms of the minimum of the action integral instead of 
the minimum energy path would be necessary and work in this direction is in progress.
It is worth to stress here again that in the HFB framework the
masses of the fragments are determined at the scission point and  not in the
fission path far from this point.

\section{Conclusions}

The analysis of the potential energy surface in a mean-field model such as the
Hatree-Fock-Bogoliubov theory with the D1S Gogny force can be successfully
applied to explain many features of cluster radioactivity.  The very mass 
asymmetric fission valley can be easily found in the potential energy surface of
the HFB theory. To determine the fission path at the bottom of this valley the
octupole moment as the collective coordinate leading to fission is required. The
results for Th isotopes  described in this paper are in reasonable good
agreement with experimental fragment's masses and half-lives.

\section*{Acknowledgments}

L.M.R. acknowledges financial support from the DGI of the MEC (Spain)
under Project FIS2004-06697 and M.W. acknowledges financial support 
by Grants No.\ FIS2005-03142 from MEC (Spain) and FEDER
and No. N202 179 31/3920 from MNiSW (Poland).

%%%%%%%%%%%%%%%%%%%%%%%%%%%%%%%%%%%%%%%%%%%%%%%

\begin{thebibliography}{0}
\bibitem{1}	A. Sandulescu, D.N. Poenaru and W. Greiner,
		{\it Sov. J. Part Nucl.} \textbf{11}, 528 (1980).
\bibitem{2}	H.J. Rose and G.A. Jones, 
		{\it Nature} \textbf{307}, 245 (1984).
\bibitem{3} 	M. Warda, J. L. Egido, L. M. Robledo and K. Pomorski,
		{\it Phys. Rev.} {\bf C 66}, 014310 (2002).
\bibitem{4} 	M. Warda,  K. Pomorski, J. L. Egido and L. M. Robledo,
		{\it Int. J. Mod. Phys.} {\bf E 14}, 403 (2005).
\bibitem{5}     M. Warda, K. Pomorski, J. L. Egido and L. M. Robledo,
		{\it J. Phys.}  {\bf G}: {\it Nucl. Part. Phys.} {\bf 31}, S1555 (2005).
\bibitem{6}	J.L. Egido and L.M.  Robledo,
 		{\it Nucl. Phys.} {\bf A 738}, 31 (2004). 
\bibitem{7}	L.M.  Robledo and M. Warda,
		in preparation.
\bibitem{8}	G. Audi, A.H. Wapstra and C. Thibault,
		{\it Nucl. Phys.} {\bf A  729}, 337 (2003).
\bibitem{9}	G. Audi, O. Bersillon, J. Blachot and A.H. Wapstra,
		{\it Nucl. Phys.} {\bf A  729}, 3 (2003).

		
\end{thebibliography}
\end{document}